\documentclass[sigconf,authorversion,natbib=true]{acmart}
\AtBeginDocument{%
  \providecommand\BibTeX{{%
    \normalfont B\kern-0.5em{\scshape i\kern-0.25em b}\kern-0.8em\TeX}}}

\usepackage{textcomp}
\usepackage[T1]{fontenc}
\usepackage[frozencache,cachedir=./pyg]{minted}
\usepackage{tabularx}
\usepackage{fontawesome}
\usepackage{multirow}

\usepackage[htt]{hyphenat}

\copyrightyear{2022}
\acmYear{2022}
\setcopyright{acmcopyright}
\acmConference[SIGIR '22]{Proceedings of the 45th International ACM SIGIR Conference on Research and Development in Information Retrieval}{July 11--15, 2022}{Madrid, Spain}
\acmBooktitle{Proceedings of the 45th International ACM SIGIR Conference on Research and Development in Information Retrieval (SIGIR '22), July 11--15, 2022, Madrid, Spain}
\acmPrice{15.00}
\acmDOI{10.1145/3477495.3531738}
\acmISBN{978-1-4503-8732-3/22/07}

\begin{document}

\fancyhead{}

\title{\texttt{ir\_metadata}: An Extensible Metadata Schema for IR Experiments}

\author{Timo Breuer}
\orcid{0000-0002-1765-2449}
\affiliation{
  \institution{TH Köln}
  \country{Germany}
}
\email{timo.breuer@th-koeln.de}

\author{J{\"u}ri Keller}
\orcid{0000-0002-9392-8646}
\affiliation{
  \institution{TH Köln}
  \country{Germany}
}
\email{jueri.keller@th-koeln.de}

\author{Philipp Schaer}
\orcid{0000-0002-8817-4632 }
\affiliation{
  \institution{TH Köln}
  \country{Germany}
}
\email{philipp.schaer@th-koeln.de}

\renewcommand{\shortauthors}{Breuer, Keller, Schaer}

\begin{abstract}
The information retrieval (IR) community has a strong tradition of making the computational artifacts and resources available for future reuse, allowing the validation of experimental results. Besides the actual test collections, the underlying run files are often hosted in data archives as part of conferences like TREC, CLEF, or NTCIR. Unfortunately, the run data itself does not provide much information about the underlying experiment. For instance, the single run file is not of much use without the context of the shared task's website or the run data archive. In other domains, like the social sciences, it is good practice to annotate research data with metadata. In this work, we introduce \texttt{ir\_metadata} - an extensible metadata schema for TREC run files based on the PRIMAD model. We propose to align the metadata annotations to PRIMAD, which considers components of computational experiments that can affect reproducibility. Furthermore, we outline important components and information that should be reported in the metadata and give evidence from the literature. To demonstrate the usefulness of these metadata annotations, we implement new features in \texttt{repro\_eval} that support the outlined metadata schema for the use case of reproducibility studies. Additionally, we curate a dataset with run files derived from experiments with different instantiations of PRIMAD components and annotate these with the corresponding metadata. In the experiments, we cover reproducibility experiments that are identified by the metadata and classified by PRIMAD. With this work, we enable IR researchers to annotate TREC run files and improve the reuse value of experimental artifacts even further. 
\end{abstract}



\maketitle

\section{Introduction}

Information retrieval (IR) research is an inherently data-driven process. Typically, IR experiments follow a common retrieval and evaluation pipeline that is based on (semi-)public datasets allowing the validation of experimental results and meta-analyses. This process would be supported by annotating the research artifacts and resources, i.e., TREC-style formatted run files, with a complementing set of metadata. While in other domains, metadata has been established for years, for instance, by the \textit{Data Documentation Initiative (DDI)} \cite{vardigan_data_2008} 
standard in the social sciences, or by the \textit{Digital Imaging and Communications in Medicine (DICOM)} \cite{graham2005dicom} standard for medical image data, the IR community has missed the opportunity to make the experimental artifacts even more valuable by annotating them with metadata. An extensive and recent overview of this topic is provided by Leipzig et al. \cite{DBLP:journals/patterns/LeipzigNHRG21}.

When taken out of context, the run data itself is not of any use. Without TREC's data archive, it is not conclusive for what kinds of experiments, e.g., for which test collection or track the run data was derived. Even though the TREC-style run format also contains a run identifier in the last column, it can be ambiguous from which experiment the run data originates.
Metadata annotations of TREC run files would provide information about the context and the experiments to which the run files belong. Furthermore, they would reduce the effort of meta-evaluations that are one of three major challenges w.r.t. reproducible IR research according to Ferro \cite{DBLP:journals/jdiq/Ferro17}. Searching metadata fields of annotated run files is a cost-efficient solution to find adequate baselines. Otherwise, these baselines can only be found by more time-consuming literature research.

Recently, a Nature survey \cite{baker20161} stimulated discussions about reproducible research and made aware of the so-called reproducibility crisis that nearly affects all scientific disciplines, according to the survey’s results. A large-scale reproducibility study in the computer sciences \cite{collberg2015repeatability} showed that, even if the code of the experiments is available, it is sometimes not enough to reproduce the original experiments. Reproducibility in the computer sciences is not only about providing the source code or other experimental artifacts, but it is also a question of documentation and communication in a wider sense.

When reproducing IR experiments and disseminating the results, authors use the terminology of \textit{reproducibility} and related terms in different and sometimes inconsistent ways. Even though the ACM policy of Artifact Review and Badging\footnote{\url{https://www.acm.org/publications/policies/artifact-review-and-badging-current}} provides definitions regarding \textit{repeatability}, \textit{reproducibility}, and \textit{replicability}, there is still enough freedom of interpretation how these definitions can be applied to the IR experiment in different ways. For instance, ECIR hosts a dedicated reproducibility track, where the authors use terms like reproducibility, replicability, robustness, and generalizability as they see fit. Judging from the terminology alone, it is unclear what exactly has been changed and under which circumstances the former experiments could be validated. PRIMAD \cite{DBLP:journals/dagstuhl-reports/FreireFR16,DBLP:journals/sigir/FerroFJKLZ16} can be seen as an answer to this underspecification. The acronym stems from the components of a typical experiment in the computational sciences, including the \textbf{P}latform, \textbf{R}esearch goal, \textbf{I}mplementation, \textbf{M}ethod, \textbf{A}ctor, and the \textbf{D}ata. Defining which PRIMAD components were modified facilitates a specification of how the reproduced experiment ``adds up'' to the former to which the experiment is compared. 

Our metadata schema is based on PRIMAD, and the metadata information is classified w.r.t. the PRIMAD components. Inspired by the idea of Open Runs \cite{DBLP:conf/ntcir/VoorheesRS16}, we want to promote the idea of making the experimental artifacts even more transparent in order to foster reproducibility. From a practical point of view, we propose to write the metadata, similar to a file header, directly into the beginning of run files (details are provided in Section \ref{sec:metadata}). The contributed resources of this work are as follows: 

\begin{itemize}
    \item We outline a \textbf{schema for metadata} annotations of TREC run files based on the PRIMAD model and review what kinds of reproducibility issues can occur w.r.t. the corresponding PRIMAD components that can be mitigated by reporting the required information in the metadata. 
    \item We introduce a \textbf{new software feature} supporting the outlined metadata schema for the evaluation toolkit \texttt{repro\_eval} used in system-oriented IR reproducibility studies.
    \item In order to demonstrate the potential of metadata annotations, we curate and provide an annotated \textbf{open-access dataset}\footnote{\url{https://zenodo.org/record/5997491}} with TREC-style formatted run data used in our experiments to showcase what kinds of meta-evaluation are made possible by our metadata schema.
\end{itemize}

Additionally, we provide a \textbf{dedicated website}\footnote{\url{https://ir-metadata.org/}} for the outlined metadata schema and the related resources. The website compiles additional documentation about the metadata and checklists that can be used as a reference when annotating run files. Furthermore, the website includes pointers to the run dataset and an interactive Jupyter notebook that can be run on Google Colab.

\section{Related Work}

Leipzig et al. review existing metadata formats for the computational sciences and evaluate how they make experiments more reproducible \cite{DBLP:journals/patterns/LeipzigNHRG21}. They distinguish between different metadata levels, including the (1) input, (2) tools, (3) statistical reports and notebooks, (4) pipelines, preservations, and binding, and (5) publication. Inspired by the rigorous descriptions of datasheets in the electronic industry, Gebru et al. highlight that there are currently no industry standards for documenting machine learning datasets \cite{DBLP:journals/cacm/GebruMVVWDC21}. They outline a catalog with 57 questions that dataset curators should address to make communication easier between dataset creators and consumers. They emphasize that their proposal is not intended to be automated since the annotation quality would benefit from careful reflection during the maintenance process. Gäde et al. propose a manifesto to make resources of Interactive Information Retrieval (IIR) studies more reusable \cite{DBLP:conf/chiir/GadeKHBP21}. In this context, \textit{resources} comprise the research design, the data, and the underlying infrastructure. Their manifesto is based on eight principles. They derive concrete actions that experimenters should consider when conducting and documenting IIR studies. With special regards to IR research, Potthast et al. introduce a taxonomy \cite{DBLP:books/sp/19/PotthastGWS19} that classifies approaches towards reproducibility into \textit{supportive}, \textit{proactive}, and \textit{reactive} actions. While supportive actions are often realized with Evaluation-as-a-Service (EaaS) platforms \cite{DBLP:journals/jdiq/HopfgartnerHMEB18} like TIRA \cite{DBLP:books/sp/19/PotthastGWS19}, we review some of the existing work for pro- and reactive actions in the following. 

\subsection{Proactive approaches}

By following the proactive approach, researchers usually prepare the experimental setup from the early beginning in order to make it reusable for others. Recently, Ferro and Kelly surveyed the IR community about badging efforts \cite{DBLP:journals/sigir/Ferro018} as a proactive effort and could show by the study's results that there is an overall positive attitude towards artifact badging that is now part of major IR venues\footnote{\url{https://sigir.org/general-information/acm-sigir-artifact-badging/}}.

Besides the outlined attempts on the organizational level, the experimenters themselves can rely on various software tools that help to prepare an experimental setup for reproducibility. For instance, it is possible to record system calls with ReproZip \cite{DBLP:conf/tapp/ChirigatiSF13}, enforce a tighter integration between publication and implementation by executable papers \cite{DBLP:journals/procedia/GorpM11,DBLP:journals/pvldb/DittrichB15,DBLP:conf/kcap/RichardsonCBSRD21}, 
or \textit{LaTeX} \cite{DBLP:journals/corr/abs-2010-01482}. Furthermore, there exist solutions for describing the method by visualizations of workflows \cite{DBLP:journals/csur/IvieT18}, by the Common Workflow Language (CWL) \cite{DBLP:journals/corr/abs-2105-07028}, with container frameworks like Popper \cite{DBLP:conf/ipps/JimenezSWMLMAA17}, or Docker in general \cite{DBLP:journals/sigops/Boettiger15}.

Lipani et al. address the need for machine-readable annotations of IR experiments by introducing nanopublications \cite{DBLP:conf/clef/LipaniPAH14}. They propose an ontology for the annotation of experimental components in reference to which the experiment should be described on a meta-level. Their ontology covers the aspects of experimental evaluation and the actual retrieval system. In a similar way, but with no special focus on IR experiments, Miksa and Rauber propose an ontology and use it to describe experimental workflows in the computational sciences with a controlled vocabulary \cite{DBLP:journals/ws/MiksaR17}.

Ram promotes the use of open-source version control systems (VCS), specifically git, to trace the progression of experimental setups in the computational sciences \cite{DBLP:journals/scfbm/Ram13}. Besides offering a collaborative platform, git (or VCS in general) contributes to better transparency, providing insights about the provenance. Fortunately, there is an increasing trend toward sharing the source code of computational experiments on GitHub as analyzed by Färber \cite{DBLP:conf/jcdl/000120}. 

Voorhees et al. introduced the idea of Open Runs as part of TREC 2015 \cite{DBLP:conf/ntcir/VoorheesRS16}, according to which a public GitHub repository backs the run submission. The authors emphasize the better reproducibility of the experiments and the availability of adequate baselines for future research. 
Approximately 25\% of the participants made their run submission \textit{open}, but, unfortunately, the authors themselves considered their first attempts as ``\textit{too simplistic}'' referring to problems like underspecified instructions or external data dependencies.

With special regards to research software, Kriegel et al. advocate ``\textit{to reuse and share common code}'' \cite{DBLP:journals/kais/KriegelSZ17}. Recently, several toolkits were introduced with a special focus on making the experiments more reproducible. Building up on the popular open-source library Lucene, the Anserini toolkit offers a specifically tailored interface for reproducible scientific experiments \cite{DBLP:journals/jdiq/YangFL18}. Pyserini was introduced for a better integration of Anserini and modern deep learning frameworks that are usually implemented with Python \cite{DBLP:conf/sigir/LinMLYPN21}. Similarly, Pyterrier is a Python interface to Terrier - another commonly used and established toolkit for IR experiments \cite{DBLP:conf/ictir/MacdonaldT20}. Co-related to Pyterrier, \texttt{ir\_datasets} is a data catalog which offers a standardized interface to commonly used IR test collections \cite{DBLP:conf/sigir/MacAvaneyYFDCG21} that allows to load and prepare data in a reproducible way.  

\subsection{Reactive approaches}

The reactive approach towards reproducibility challenges the reproducers to reimplement the experimental setup when little or no artifacts of the original experiment are available. Starting in 2015, the ECIR Reproducibility Track offers the possibility to report about reproduced experiments independent of success \cite{DBLP:conf/ecir/2015}. Most recently, also SIGIR included a reproducibility track as part of the call for papers. Similarly, the cross-venue workshop CENTRE invited its participants to validate their reimplementations of previous work submitted to TREC, CLEF, or NTCIR \cite{DBLP:conf/clef/FerroMSS18a,DBLP:conf/clef/FerroFMSS19}.

Fundamental to the reactive reproducibility analysis is the availability of the run files that are usually hosted in archives as part of shared task initiatives at TREC, CLEF, or NTCIR. TREC's website\footnote{\url{https://trec.nist.gov/}} provides access to a password-protected area where experimenters have insights into the results that were submitted to previous tracks. Armstrong et al. developed a central web-based run archive called EvaluatIR that could be consulted when comparing new methods to previously uploaded runs \cite{DBLP:conf/sigir/ArmstrongMWZ09a}. Once uploaded, the run could be shared by a permanent URL and could be validated in reference to other baselines. Another more recent public web-based service was the RISE platform developed by Yang and Fang \cite{DBLP:conf/ictir/Yang016}. RISE made it possible to validate 20 retrieval methods with 16 TREC test collections on a shared technical platform. Besides the already implemented methods, RISE offered the possibility to integrate new experiments for registered users. However, both services are currently not available online.

The DIRECT system is an IR evaluation infrastructure that ``\textit{supports the archiving, access, citation, dissemination, and sharing of the experimental results}'' \cite{DBLP:conf/ercimdl/NunzioF05,DBLP:conf/clef/AgostiNF06}. The entire infrastructure is based on eight conceptual areas, including (1) the evaluation activity, (2) the experiment, (3) the resource management, (4) the experimental collection, (5) the measurement, (6) the metadata, (7) visual analytics, and (8) bibliographical data. Although PRIMAD has a different categorization on the conceptual level, all of DIRECT's aspects are part of the introduced metadata schema in Section \ref{sec:metadata} as well. DIRECT's RESTful web service allows the structured querying of experimental artifacts. In comparison, directly writing the metadata into the run files, introduces a more lightweight approach for structured comparisons of experimental rankings.

More recently, Breuer et al. introduced a framework with reproducibility measures that can be used to validate reimplementations when a reference ranking of the original implementation is available \cite{DBLP:conf/sigir/Breuer0FMSSS20}. The corresponding reproducibility measures are also available in an open-source software toolkit \cite{DBLP:conf/ecir/BreuerFMS21}.

\section{Metadata Annotations}
\label{sec:metadata}
According to ISO 23081-1, the metadata schema is defined as a \textit{``logical plan showing the relationships between metadata elements''}. Similar to other metadata schemas and protocols \cite{DBLP:conf/clef/AgostiNF06,DBLP:conf/chiir/GadeKHBP21,DBLP:journals/patterns/LeipzigNHRG21}, our proposed schema is based on related conceptual components as well. More specifically, we define the metadata annotations based on the components of the PRIMAD model. To the best of our knowledge, the PRIMAD taxonomy has not been put into practice yet. Even though PRIMAD is well-grounded and considers key elements that might affect the reproducibility of experimental outcomes, it is a rather theoretical concept leaving several subcomponents underspecified. Yet, we think these abstract definitions allow enough flexibility to report details as required for reproducible experimentation. We argue that it does not seem reasonable to follow a strict annotation schema as some details do undeniably not affect the reproducibility, e.g., reporting a GPU model when it is not used in the experiments. On the other hand, it is simply not feasible to think of all subcomponents, which will be crucial for reproducible experimentation in the future. For this reason, we introduce a metadata schema inspired by PRIMAD for which we propose a set of essential subcomponents that should be reported if feasible. In the following, we outline these for each PRIMAD component. While PRIMAD has originally been described in two different ways, covering system- and user-oriented experiments separately, we mostly focus on the system-oriented definitions relevant to IR experiments. 

From a practical point of view, we propose to write the metadata, similar to a file header, directly into the beginning of run files. Recently, the \texttt{trec\_eval} toolkit introduced the support of comments in run and qrels files\footnote{\url{https://github.com/usnistgov/trec_eval/issues/20}}. In order to allow the aforementioned flexibility of PRIMAD, we propose to add the metadata in the comments with YAML syntax. Besides good readability, it can be easily extended as required in future IR experiments. For more specific details about the YAML formatting and the encodings that are required by ISO 23081-1, we refer the reader to our project website that covers checklists including \textit{descriptions}, \textit{encodings}, and \textit{YAML types} for each metadata field. The website is easy to maintain, and we aim at developing it in a collaborative process with the community in the future. Its source code is publicly hosted on GitHub and can be easily extended by pull requests, for instance, when the checklists need updates or if it is required to adapt the terminology for certain descriptors. In the following, we outline what kinds of details for each PRIMAD component should be considered when annotating run files for reproducibility. Therefore, we provide anecdotal evidence from the existing body of literature and derive the required metadata fields from the corresponding studies.

\subsection{Platform}

The Platform comprises the hard- and software underlying the actual Implementation \cite{DBLP:journals/sigir/FerroFJKLZ16}. Besides reporting details on the hard- and software, the operating systems and the corresponding system kernel should be considered as well. Lin and Zhang reproduced the experiments of the entire Open-Source IR Reproducibility Challenge after four years~\cite{DBLP:conf/ecir/LinZ20}. When running the old experiments on newer hardware with a more modern operating system, they could reproduce the same results for only one out of seven retrieval systems. The remaining systems resulted in exceptions, segmentation faults, and compilation errors or depended on external resources that were not available anymore. In order to provide a bare minimum of the Platform, we propose to report the subcomponents given in Figure~\ref{fig:platform}. Basically, information about the hardware, the operating system, and the software should be documented, whereas we distinguish between the \textit{software} subcomponent and the actual Implementation. The Platform describes all layers below the Implementation, and as such, it also includes software libraries on which the experiment's Implementation builds up.

\begin{figure}
\begin{minted}[breaklines,fontsize=\small]{yaml}
platform:
  hardware:
    cpu:
      model: 'Intel Xeon Gold 6144 CPU @ 3.50GHz'
      architecture: 'x86_64'
      operation mode: '64-bit'
      number of cores: 16
    ram: '64 GB'
  operating system:
    kernel: '5.4.0-90-generic'
    distribution: 'Ubuntu 20.04.3 LTS'
  software:
    libraries: 
      python:
        - 'scikit-learn==0.20.1'
        - 'numpy==1.15.4'
      java:
        - 'lucene==7.6'
    retrieval toolkit: 
      - 'anserini==0.3.0'
\end{minted}
\caption{Platform metadata example}
\label{fig:platform}
\end{figure}

\subsection{Research goal} The Research goal describes the purpose of the study. If the experiment is aligned to the Cranfield paradigm, as is often the case in IR experiments, the research goal is a high-quality ranking \cite{DBLP:journals/sigir/FerroFJKLZ16}. But in general, the study can focus on other aspects that are based on research questions and the corresponding hypothesis. 

As pointed out by Fuhr \cite{DBLP:journals/sigir/Fuhr17,DBLP:journals/sigir/Fuhr20}, it is crucial for the validity to formulate hypotheses before the experiment is conducted, referring to the multiple comparison problem \cite{DBLP:journals/sigir/Fuhr20}. Likewise, drawing conclusions from results with underpowered statistics \cite{DBLP:conf/kdd/KohaviDFLWX12} and effect sizes \cite{DBLP:journals/sigir/Fuhr17}, or relying on overused test collections \cite{DBLP:conf/sigir/Carterette15} may lead to invalid and thus not reproducible outcomes. The meta-evaluations by Armstrong et al. demonstrate the stagnating overall progress when there are flaws in the research design \cite{DBLP:conf/cikm/ArmstrongMWZ09}, i.e., choosing weak baselines that are still an issue after an entire decade \cite{DBLP:conf/sigir/YangLYL19}. More recently, it was shown by Ferrante et al. that evaluation measures should be interval-scaled in order to conduct statistical significance tests with valid outcomes \cite{DBLP:journals/access/FerranteFF21}.

Besides other PRIMAD components, the Research goal is usually reported in the publication that is published at a venue. It should be reported in the metadata by identifiers as provided by arXiv.org, dblp.org, or the DOI. If available, an abstract and a short description of the experimental evaluations should be reported, including the evaluation measures, the baselines to that it is compared, and the significance test with the correction method. We propose to implement the metadata of the Research goal as illustrated in Figure~\ref{fig:research_goal}. The baseline methods are documented as a sequence of YAML scalars, if the experiment is compared to multiple runs. In this case, the Actor is an ``experimenter'' who conducts a new experiment and evaluates a new retrieval method to reasonable baseline methods. If the experiment is reproduced, the Actor is documented in the metadata as a ``reproducer'' and a single run tag, which corresponds to the target of the reproduction, should be reported in the baseline field. We assume that the documented significance tests apply to all ``reported measures'' in the publication and likewise to all documented baselines. Regarding meaningful significance tests for specific IR measures, we refer the reader to the already mentioned work by Ferrante et al. \cite{DBLP:journals/access/FerranteFF21}.

\begin{figure}
\begin{minted}[breaklines,fontsize=\small]{yaml}
research goal:
  venue:
    name: 'SIGIR'
    year: '2020'
  publication:
    dblp: 'https://dblp.org/rec/conf/sigir/author'
    arxiv: 'https://arxiv.org/abs/2010.13447'
    doi: 'https://doi.org/10.1145/3397271.3401036'
    abstract: 'In this work, we analyze ...'
  evaluation:
    reported measures: 
      - 'ndcg'
      - 'map'
      - 'P_10'
    baseline: 
      - 'tfidf.terrier'  
      - 'qld.indri'
    significance test: 
      - name: 't-test' 
        correction method: 'bonferroni'
\end{minted}
\caption{Research goal metadata example}
\label{fig:research_goal}
\end{figure}

\subsection{Implementation}
As described by Ferro et al. \cite{DBLP:journals/sigir/FerroFJKLZ16}, the Implementation is closely related to the Method. What is more formally described by the Method is translated by the Implementation into operations that can be conducted \textit{in silico}. As part of the aforementioned large-scale reproducibility study, the code repositories of approximately 600 ACM papers were analyzed \cite{collberg2015repeatability} with rather disillusioning results. Even if the code is made publicly available, it does not guarantee reproducibility. In this sense, the code is a different means to communicate the experiments, and the Actor implementing it should take care of writing clean and understandable code and annotate it with meaningful comments where necessary.

Different implementations of the same retrieval method can yield deviating results \cite{DBLP:conf/sigir/MuhleisenSLV14,DBLP:journals/kais/KriegelSZ17,DBLP:conf/ecir/KamphuisVBL20}. With special regard to the runtime evaluation, Kriegel et al. demonstrated that different Implementations of the same Method could result in runtimes that differ by several orders of magnitude \cite{DBLP:journals/kais/KriegelSZ17}. As shown by Kamphuis et al. \cite{DBLP:conf/ecir/KamphuisVBL20} in a dedicated reproducibility study, the reimplemented BM25 methods do not yield significantly different results, but still, the absolute scores are different. While this might not affect the reproducibility of the overall system performance, it is still an open question of how inconsistent rankings affect user-oriented experiments.

Even though our work favors the idea of Open Runs for which the source code should be publicly available, e.g., on GitHub, transparency is not a hard requirement for reproducibility. For instance, closed source software might be part of the experiments, and it could still be made available as binaries or at least documented through command line calls \cite{DBLP:journals/csur/IvieT18} in the metadata.

We propose including the Implementation in the metadata by referring to the public git repository and including the specific commit or release version if available. The command line calls should document the concrete execution of the computations. In sum, we propose the metadata example given in Figure~\ref{fig:implementation}.

\begin{figure}
\begin{minted}[breaklines,fontsize=\small]{yaml}
implementation:
  executable:
    cmd: './bin/search arg01 arg02 input output'
  source:
    lang: 
      - 'python' 
      - 'c'
    repository: 'github.com/castorini/anserini'
    commit: '9548cd6'
\end{minted}
\caption{Implementation metadata example}
\label{fig:implementation}
\end{figure}

\subsection{Method}
In IR research, the study's focus lies on the actual retrieval approach that is covered by the Method in the PRIMAD model \cite{DBLP:journals/sigir/FerroFJKLZ16}. It describes the mapping of query-document pairs to a ranking score. 

Underspecified or missing details about processing steps or parameters can obstruct reproducible results. Although there exist solutions that integrate the research descriptions with the experimental implementations more tightly \cite{DBLP:journals/procedia/GorpM11}, most of the scientific results are usually communicated through journals or conference proceedings, which are not the perfect medium to report technical details. Furthermore, when implementing the Method, compromises between the theory and practice must be made. For instance, the BM25 implementation of Lucene strictly follows the original formulation by Robertson et al. but represents the document length by a one-byte value that only allows 256 different lengths \cite{DBLP:conf/ecir/KamphuisVBL20}. When using software libraries, much effort is required to trace deviations from the ideal theoretical Method, and it is almost infeasible to know all compromises the Implementation implies.

Lin and Yang showed that score ties have to be broken deterministically; otherwise, they can affect the reproducibility of the document rankings \cite{DBLP:conf/sigir/LinY19}. To achieve this, they favor an external collection with document identifiers. When indexing documents, preprocessing plays an important role. As shown by Roy et al. \cite{DBLP:journals/jdiq/RoyMG18}, the removal of markup artifacts can have an impact on reproducibility. Similarly, Ferro and Silvello demonstrate the influence of preprocessing operations, including stemmer, tokenizers, and stopword lists by a component-wise analysis \cite{DBLP:conf/sigir/FerroS16}. An example of metadata annotations for the Method is given in Figure~\ref{fig:method}. Since more than one retrieval method could be used for the final ranking, the metadata schema uses YAML \textit{mappings}. As shown, the schema also allows documenting multi-stage ranking architectures by referring to the name of the previous retrieval method of which the output is reranked. Likewise, it is possible to document other pipeline stages like an interpolation between the scores of two or more retrieval methods.

\begin{figure}
\begin{minted}[breaklines,fontsize=\small]{yaml}
method:
  automatic: 'true'
  score ties: 'reverse alphabetical order'
  indexing:
    tokenizer: 'lucene.StandardTokenizer' 
    stemmer: 'lucene.PorterStemFilter' 
    stopwords: 'lucene.StandardAnalyzer' 
  retrieval:
    - name: 'bm25' 
      method: 'lucene.BM25Similarity'
      b: 0.4
      k1: 0.9
    - name: 'lr reranker'
      method: 'sklearn.LogisticRegression' 
      reranks: 'bm25' 
    - name: 'interpolation' 
      weight: 0.6 
      interpolates: 
        - 'lr reranker' 
        - 'bm25' 
\end{minted}
\caption{Method metadata example}
\label{fig:method}
\end{figure}

\subsection{Actor}

The Actor component represents the experimenter who conducts the experiments \cite{DBLP:journals/sigir/FerroFJKLZ16}. It is the one who operates the computer, implements the experiments, types commands, et cetera. The Actor is obliged to avoid a flawed experimental setup \cite{DBLP:books/sp/19/Fuhr19} and should take care of scientific rigor.

According to Ivie and Thain most issues of reproducibility arise from finding compromises between the concrete instructions a computing machine requires and the more abstract means of human communications in computational research \cite{DBLP:journals/csur/IvieT18}. When repeating experiments, the Actors has the role of a reproducer (A') who can also be the future self of the original Actor.

Even though the ultimate goal should be Actor-independency like it is made possible with EaaS platforms \cite{DBLP:journals/jdiq/HopfgartnerHMEB18,DBLP:books/sp/19/PotthastGWS19}, it might be reasonable to include Actor profiles in the metadata. The actual Actor is often hidden behind the co-authorship of a publication, and we favor making the Actor more explicit. An ORCID is a unique identifier that is assigned to researchers. As part of a shared task, the participants often assign themselves team names, e.g., the research group's name. Reporting the research domain of the researcher provides information on which expertise is required to reproduce the experiments. If available, it is helpful to report e-mail addresses and other contacts like social media profiles (cf. Figure~\ref{fig:actor}).

\begin{figure}        
\begin{minted}[breaklines,fontsize=\small]{yaml}
actor:
  name: 'Jimmy Lin' 
  orcid: '0000-0002-0661-7189'
  team: 'h2oloo' 
  fields: 
    - 'nlp'
    - 'ir'
    - 'databases'
    - 'large-scale distributed algorithms'
    - 'data analytics'
  mail: 'jimmylin@uwaterloo.ca' 
  role: 'experimenter' # or 'reproducer'
  degree: 'Ph.D.' 
  github: 'https://github.com/lintool' 
  twitter: 'https://twitter.com/lintool' 
\end{minted}
\caption{Actor metadata example}
\label{fig:actor}
\end{figure}

\subsection{Data}

By its original definition, the Data component comprises the input data, and the parameters required to run the experiments \cite{DBLP:journals/sigir/FerroFJKLZ16}. In this regard, our metadata slightly deviates from the definitions by Ferro et al. as we propose to report the parameters as part of the Implementation.

According to Jones et al. \cite{DBLP:conf/cikm/JonesTMSS14} and Fuhr \cite{DBLP:journals/sigir/Fuhr17}, it is not always assessable what characterizes a test collection, how it compares to other collections, and conclude why specific methods perform well. A recent analysis has shown that test collections may not be suitable for every system type, i.e., evaluating neural retrieval approaches based on document pools drawn from results of mostly keyword-based retrieval methods may result in evaluation bias \cite{DBLP:conf/sigir/YilmazCMC20}. On a more practical level, Ivie and Thain criticize the separation of code and data \cite{DBLP:journals/csur/IvieT18}. Likewise, the use of private or sensitive \cite{DBLP:journals/csur/IvieT18} as well as pay-walled \cite{DBLP:conf/clef/FerroFMSS19} data collections can hinder reproducers from reimplementing the experiments.

The metadata information about the test collection should be reported by its name, the source from which it can be retrieved, and the location of topics and qrels files. The recently introduced dataset catalog \texttt{ir\_datasets} \cite{DBLP:conf/sigir/MacAvaneyYFDCG21} offers an interface to a vast amount of standard test collection, and its identifiers should be included in order to make the test collection more explicit. IR experimenters often cross-validate retrieval models on different data folds during training or parameterizing retrieval methods. These should be explicitly reported in the metadata and we recommend practitioners to follow the naming conventions of \texttt{ir\_datasets} (if available). Besides the actual test collection, external data is often involved in the experiments. These include but are not limited to thesauri, word embeddings, learned weights of deep learning models, and others (cf. Figure~\ref{fig:data}).

\begin{figure}
\begin{minted}[breaklines,fontsize=\small]{yaml}
data:
  test_collection:
    name: 'The New York Times Annotated Corpus'
    source: 'catalog.ldc.upenn.edu/LDC2008T19'
    qrels: 'trec.nist.gov/data/core/qrels.txt'
    topics: 'trec.nist.gov/data/core/core_nist.txt'
    ir_datasets: 'nyt/trec-core-2017' 
  training_data: 
    - name: 'TREC Robust 2004'
      folds: 
        - 'disks45/nocr/trec-robust-2004/fold1'
        - 'disks45/nocr/trec-robust-2004/fold2' 
  other:
    - name: 'GloVe embeddings'
      source: 'https://nlp.stanford.edu/projects/glove/'
\end{minted}
\caption{Data metadata example}
\label{fig:data}
\end{figure}

\section{Metadata support of \texttt{repro\_eval}}

In this section, we introduce the new metadata support of \texttt{repro\_eval} that is a dedicated reproducibility evaluation toolkit featuring bindings to \texttt{trec\_eval} \cite{DBLP:conf/ecir/BreuerFMS21}. The introduced metadata schema is supported by \texttt{repro\_eval==0.4.0} in two different ways. First, we have implemented a \texttt{MetadataHandler} that, on the one hand, reads the metadata from annotated run files and, on the other hand, semi-automatically annotates run files if provided with a minimal set of the required information. Second, we introduce the analysis of annotated run files by the \texttt{MetadataAnalyzer} and the \texttt{PrimadExperiment} that, in combination, analyze the metadata information and align the reproducibility evaluations to the PRIMAD model. Depending on the deviating PRIMAD components, different reproducibility measures are part of the evaluations, and the \texttt{MetadataAnalyzer} identifies reasonable evaluations. In the following, we provide details about both features and exemplify their utility later on in Section~\ref{sec:meta-eval}. More technical details about the implementation of the \texttt{metadata} module can be found in the public repository of \texttt{repro\_eval}\footnote{\faicon{github} \url{https://github.com/irgroup/repro_eval}}.

\subsection{Automatic Annotations}
\begin{figure}
\begin{minted}[linenos,breaklines,fontsize=\small]{python}
from repro_eval.metadata import MetadataHandler

run_path='./run.txt', 
metadata_path='./metadata.yaml'
metadata_handler = MetadataHandler(run_path, metadata_path)
metadata_handler.write_metadata()
\end{minted}
\caption{Annotating runs with the \texttt{MetadataHandler}.}
\label{fig:metadata_handler}
\end{figure}

In order to lower the manual annotation effort, the \texttt{MetadataHandler} can be used to annotate runs. Given a run file, it automatically compiles the available information and appends it to the metadata header of a run file. Figure \ref{fig:metadata_handler} exemplifies how the metadata can be added to the run file in Python. When writing metadata to run files, the \texttt{MetadataHandler} fetches information regarding the Platform by reading out, for instance, the information about the CPU. Likewise, if not specified otherwise, it assumes the run file to be in the root directory of a git repository from which some of the information regarding the Implementation can be extracted. 
It is impossible for some of the PRIMAD components to determine the required information automatically. Therefore, the \texttt{MetadataHandler} has to be provided with a template YAML file (\texttt{metadata.yaml}), in which the corresponding metadata should be added manually. Specifically, some of the information about the Research Goal, the Method, or the Actor cannot be retrieved automatically yet and must be added by hand. Even though the entire metadata cannot be extracted automatically, the \texttt{MetadataHandler} reduces the manual annotation effort, which contributes to a community-wide adoption.

\subsection{Analysis of Annotations}
\begin{figure}
\begin{minted}[linenos,breaklines,fontsize=\small]{python}
from repro_eval.metadata import MetadataAnalyzer, PrimadExperiment

run_path='./run.txt'
dir_path ='./runs/'
metadata_analyzer = MetadataAnalyzer(run_path)
experiments =  metadata_analyzer.analyze_directory(dir_path)
run_candidates = experiments.get('priMad')
primad_experiment = PrimadExperiment(primad='priMad',
                      rep_base=run_candidates,...)
primad_experiment.evaluate()
\end{minted}
\caption{Analyzing runs with the \texttt{MetadataAnalyzer}.}
\label{fig:metadata_analyzer}
\end{figure}

Given two or more run files with metadata annotations, the \texttt{MetadataAnalyzer} identifies similar PRIMAD components in the metadata and proposes several reasonable reproducibility evaluations. The code snippet in Figure \ref{fig:metadata_analyzer} illustrates how the \texttt{MetadataAnalyzer} can be used to scan an entire directory with annotated run files and automatically validates the type of reproducibility experiments. After it has been initialized with a reference run, the metadata of all annotated runs in the directory is compared to that of the reference metadata, and as a result, a list containing PRIMAD experiments and the corresponding run candidates is returned. In our implementations, we distinguish the different experiment types by lower- and uppercase letters, e.g., parameter sweeps would result in \texttt{``priMad''} with the uppercase letter \texttt{M} that signifies the changes of the Method. Provided with the experiment type and the reproduced runs, the \texttt{PrimadExperiment} evaluates the experiments and the corresponding runs with the help of the \texttt{repro\_eval} measures. The reproducibility toolkit \texttt{repro\_eval} follows the naming conventions as introduced by the ACM policy on Artifact Review and Badging. More specifically, the library distinguishes between reproducibility and replicability. Reproducibility describes another group of researchers reusing the original experimental setup, which complies with reusing the test collection of the original retrieval experiments. In comparison, replicability describes another group of researchers using a different experimental setup, which complies with evaluating the reimplemented retrieval experiment with another test collection. Specific reproducibility measures can be determined depending on what kind of test collection is used to evaluate the reimplementations. For instance, if another collection than in the original experiment is used, only some of the measures can be determined. This means that the evaluations depend on the type of the PRIMAD experiment from which the reproduced runs originate. For instance, if the reproduced runs result from parameter sweeps, only the Method in the metadata changes (the test collection is the same as in the original experiment), and all reproducibility measures can be determined.  

\section{Run dataset based on Cross-Collection Relevance Feedback}

\begin{figure}
    \centering
    \includegraphics[width=0.625\linewidth]{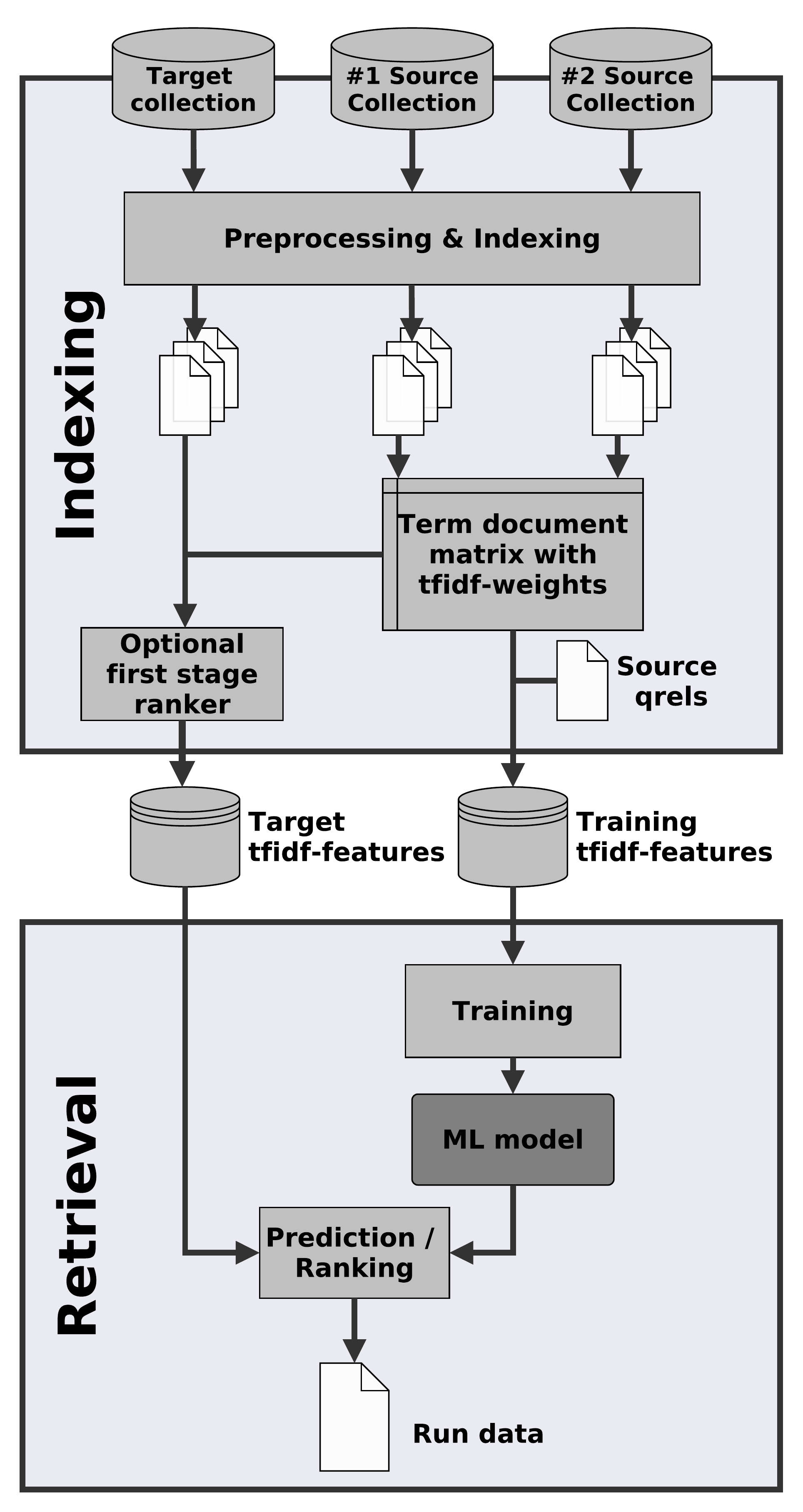}
    \caption{Cross-collection relevance feedback \cite{DBLP:conf/trec/GrossmanC17}}
    \label{fig:ccrf}
\end{figure}

To demonstrate the potential of the introduced metadata schema, we showcase its applicability by annotating a dataset of run files based on \textit{cross-collection relevance feedback} (CCRF). This retrieval method recently gained interest in the IR community, especially as part of TREC Common Core where existing topics were reused for a new dataset. The general workflow is illustrated in Figure \ref{fig:ccrf}. Dating back to 2017, Grossmann's and Cormack's (GC) approach~\cite{DBLP:conf/trec/GrossmanC17} inspired several follow-up studies. Reasons for the increased interest in this retrieval method are its simplicity and its effectiveness, being the most effective automatic submission as part of TREC Common Core 2017 (Core17)~\cite{DBLP:conf/trec/AllanHKLGV17}. 

Yu et al. (YXL) reproduced the approach by embedding it into a multi-stage ranking pipeline~\cite{DBLP:conf/ecir/YuXL19} and documenting it in the Anserini toolkit. Breuer et al. (BFFMSSS) reimplemented the approach as part of a dedicated reproducibility analysis~\cite{DBLP:conf/clef/BreuerS19} and later used that method to derive a large dataset 
of runs simulating a researcher trying to reproduce the relevance transfer method~\cite{DBLP:conf/sigir/Breuer0FMSSS20}. As part of the TREC Common Core reiteration in 2018, YXL reused the same method with another dataset~\cite{DBLP:conf/trec/YuXL18}, whereas GC themselves also applied a modified version of the method in TREC Common Core 2018~\cite{DBLP:conf/trec/GrossmanC18} that was later reproduced by Breuer et al. (BPS)~\cite{DBLP:conf/clef/BreuerPS21}. Table \ref{tab:routing_runs} provides an overview of the run files and their underlying combinations (of retrieval methods and test collections) included in our dataset. 

\begin{table}[]
\caption{Overview of the run dataset}
\begin{tabularx}{\columnwidth}{l|l|l|r}
 \toprule
 Researchers & Method & Target collections & Runs \\
 \hline
 GC~\cite{DBLP:conf/trec/GrossmanC17} & \multirow{3}{*}{GC~\cite{DBLP:conf/trec/GrossmanC17}} & Core17 & 2 \\
 YXL~\cite{DBLP:conf/trec/YuXL18,DBLP:conf/ecir/YuXL19} &  & Robust04/05, Core17/18 & 327 \\
 BFFMSSS~\cite{DBLP:conf/clef/BreuerS19,DBLP:conf/sigir/Breuer0FMSSS20} &  & Core17 & 100 \\
 \hline
 GC~\cite{DBLP:conf/trec/GrossmanC18} & \multirow{2}{*}{GC~\cite{DBLP:conf/trec/GrossmanC18}} & Core18 & 2 \\ 
 BPS~\cite{DBLP:conf/clef/BreuerPS21} &  & Robust04/05, Core17/18 & 32 \\
 \bottomrule
\end{tabularx}
\label{tab:routing_runs}
\end{table}

The general workflow of CCRF is illustrated in Figure \ref{fig:ccrf}. The underlying retrieval method follows a point-wise learning-to-rank approach where each document is assigned a probability of being relevant \cite{DBLP:journals/ftir/Liu09}. CCRF is only possible if there is an overlap of topics in the target and source test collections. For each topic, a relevance classifier is trained with the help of the relevance labels (qrels) and tifdf-features of relevant and non-relevant documents derived from a term-document matrix based on the source collection's vocabulary. The documents of the target collection are represented as tfidf-features that are also derived from the source collection's term-document matrix. The topic-specific relevance classifier assigns a relevance probability to each tfidf-feature of the target collection's documents that result in the final ranking. 

GC introduced the outlined approach at TREC Common Core 2017 \cite{DBLP:conf/trec/GrossmanC17} using only Robust04 or a combination of both Robust04 and Robust05 as the source collection(s) to rank documents of the Annotated New York Times Corpus. Even though the approach is simple and straightforward, it was the most effective automatic run submission at Core17, ranking third behind two manual runs that were slightly more effective. 

YXL~\cite{DBLP:conf/ecir/YuXL19} reproduced the approach by introducing a first-stage ranker with a keyword-based method. Instead of ranking each document of the entire target collection with the topic-specific classifier, only a list of the first 10,000 documents retrieved by the keyword-based method is reranked. BFFMSSS~\cite{DBLP:conf/clef/BreuerS19,DBLP:conf/sigir/Breuer0FMSSS20} reproduced the workflow as accurate as possible. As part of TREC Common Core 2018, YXL reused their reimplementation and submitted runs derived from the Washington Post Corpus v2. GC also submitted runs to Core18 with a slightly modified workflow. Instead of using TREC test collections as the source collections, they scraped results from search engine result pages to train topic-specific relevance classifiers. This approach is robust as shown in the regression experiments by BPS \cite{DBLP:conf/clef/BreuerPS21}.

While some of the runs were available from existing data archives, others were derived by us with Anserini's runbook\footnote{\url{https://github.com/castorini/anserini/blob/master/docs/runbook-ecir2019-ccrf.md}} that belongs to the implementations of the reproducibility analysis by Yu et al.~\cite{DBLP:conf/ecir/YuXL19}. More specifically we used Java \texttt{v8}, Lucene \texttt{v7.6}, and Anserini \texttt{v0.3.0} at commit \texttt{9548cd6b}, which were also reported in the corresponding paper, to rerun the instructions of the runbook successfully on all four test collections. All of the runs were annotated by us as far as the respective information was publicly available. The annotated run data and the corresponding metadata in separate YAML files are hosted in the aforementioned open-access data archive.

\section{Supporting reproducibility evaluations by metadata analysis}
\label{sec:meta-eval}
In this Section, we evaluate some of the annotated runs in different experimental settings. Given the metadata annotations, we align the experiments to PRIMAD. While we keep certain components fixed, others are modified to gain new insights. In the first experiment, only the Method component is varied by principled parameter changes - a setting that complies with parameter sweeps as they are usually done in computational experiments. In the second experiment, we evaluate the reproducibility of the CCRF method in reference to the original submission made by GC and the corresponding reimplementations, which translate into keeping the Data fixed, while other PRIMAD components are varied. Finally, we evaluate the generalizability in the third experiment by varying all of the PRIMAD components. All of the experimental evaluations can be rerun with the help of the Colab notebook. 

\subsection{PRIM'AD: Parameter sweeps of the Method}
Having implemented a retrieval method, the Actors usually improve the retrieval performance by finding optimal parametrizations. Sometimes, this can be realized with a systematic parameter analysis by tuning the implementations with grid search techniques. 

\begin{figure}
    \centering
    \includegraphics[width=0.49\columnwidth]{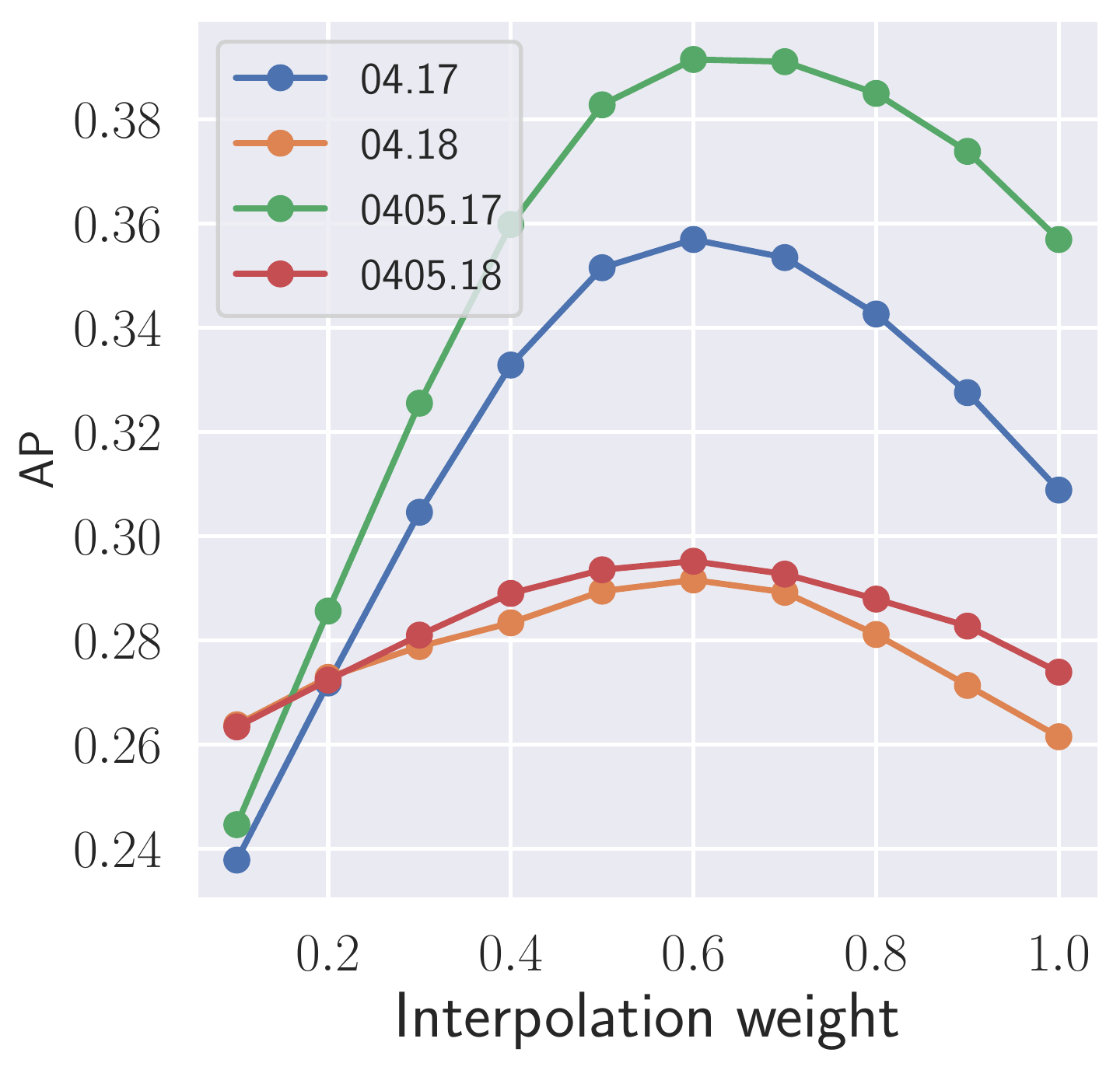}
    \includegraphics[width=0.49\columnwidth]{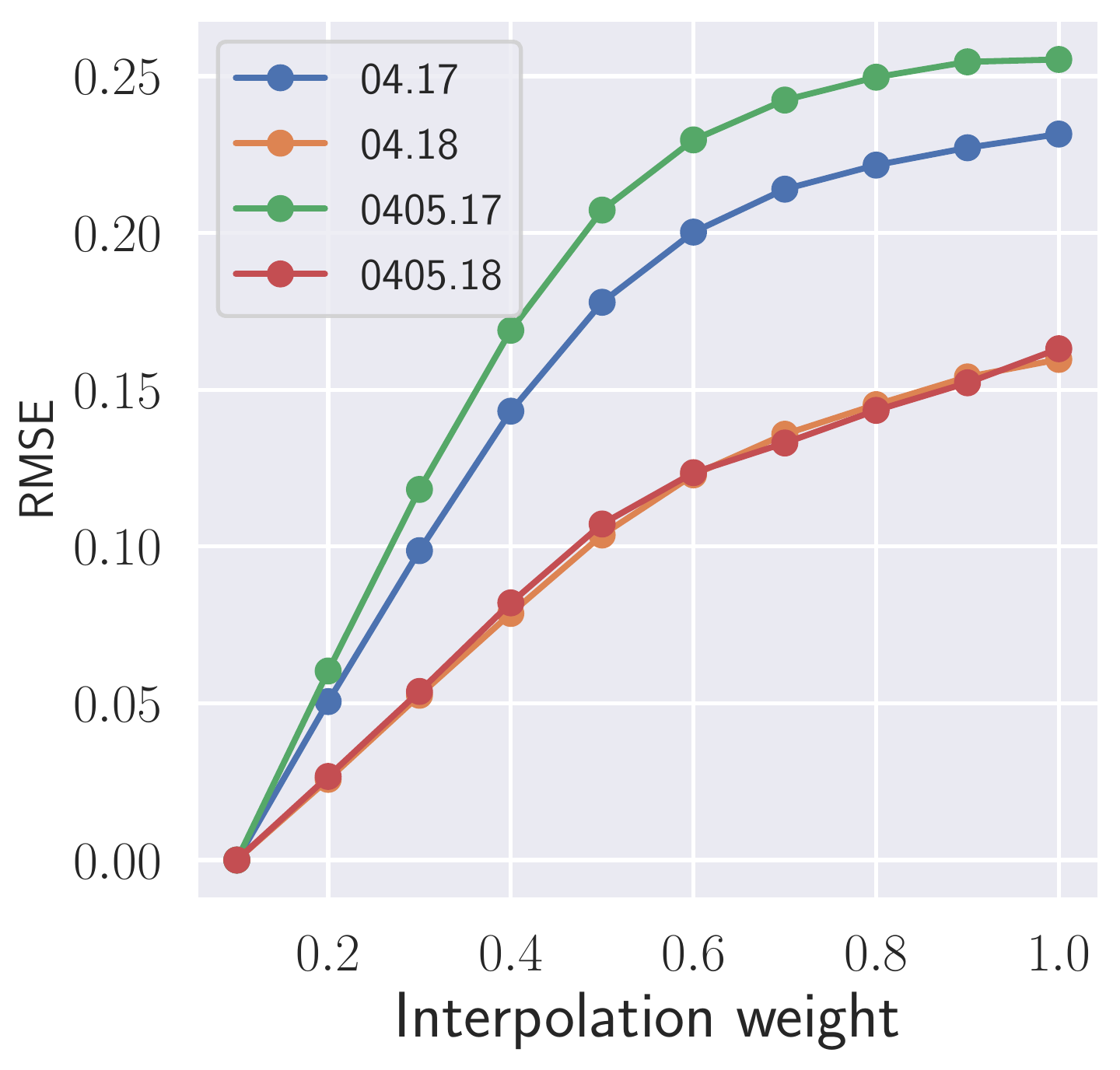}
    \caption{Parameter sweeps (of interpolation weights) evaluated by Average Precision (AP) and the corresponding error RMSE based on reimplementations by Yu et al. \cite{DBLP:conf/ecir/YuXL19}}
    \label{fig:priMad_nrmse_rbo}
\end{figure}

In this experiment, we make use of the reimplementations by YXL~\cite{DBLP:conf/ecir/YuXL19}. In contrast to the original experiments, they introduced a multi-stage ranking pipeline to the CCRF method by retrieving the first ranking with BM25 and expanded methods (including RM3 and axiomatic reranking). The initially retrieved list is then reranked by a machine learning (ML) classifier, and the scores of both the first-stage ranking and the ML reranking are interpolated with parameterizable weights. Figure \ref{fig:priMad_nrmse_rbo} evaluates the rerankings with different interpolation weights by the Average Precision (AP) and the corresponding Root-Mean-Square-Error (RMSE\textsubscript{AP}).

The evaluated runs are either made with only one source collection (Robust04) as training data and derived from two different target collections (Core17 and Core18), denoted as \texttt{04.17} and \texttt{04.18} or with two source collections as training data (Robust04 and Robust05), denoted as \texttt{0405.17} and \texttt{0405.18}. As the overall retrieval performance in terms of AP shows, there is a sweet spot around an interpolation weight of $0.6$ which indicates that the runs of our regression tests reproduce those of YXL~\cite{DBLP:conf/ecir/YuXL19}. 

The RMSE was introduced as a measure that quantifies the error between the topic score distributions of two runs \cite{DBLP:conf/sigir/Breuer0FMSSS20}. In our evaluations, it is determined between the distributions of topic scores from the reranked runs with the lowest interpolation weight ($0.1$) in comparison to the other reranked runs. While the absolute retrieval performance decreases after it peaks around an interpolation weight of $0.6$, the RMSE\textsubscript{AP} monotonically increases, meaning that the topic score distributions more and more diverge, which can be attributed to the increasing influence of the ML reranker. 

It can clearly be seen that the absolute AP scores, as well as the RMSE\textsubscript{AP} scores, differ depending on the dataset (e.g., \texttt{04.17} vs. \texttt{04.18}), which indicates that the CCRF method has a data dependency w.r.t. the combination of the source and target collections.

\subsection{P'R'I'M'A'D: Reproducing the experiments}

Reactive actions towards reproducibility \cite{DBLP:books/sp/19/PotthastGWS19} can be realized in the form of a reimplementation study where the experimental setup is repeated based on the descriptions in the publication of the original experiment. If the corresponding outputs of the original experiments are available, as is the case for TREC runs, we can use these artifacts as points of reference to which we compare the reimplementations' outputs. 

In the following experiment, we evaluate the reproducibility of the CCRF method by comparing the reimplementations of YXL and BFFMSSS to the original results by GC. Table \ref{tab:PRImAd} shows an evaluation of the reproduction quality based on the measures that were introduced by Breuer et al.~\cite{DBLP:conf/sigir/Breuer0FMSSS20}. In the following, we describe the reproductions by these measures and shortly explain them, but we refer the reader to the corresponding publications for more detailed explanations.

\begin{table}[]
\centering
\caption{Reproducibility evaluation of Grossmann and Cormack~\cite{DBLP:conf/trec/GrossmanC17} compared to Yu et al.~\cite{DBLP:conf/ecir/YuXL19} and Breuer et al.~\cite{DBLP:conf/sigir/Breuer0FMSSS20}}
\label{tab:PRImAd}
\begin{tabularx}{\columnwidth}{X|c|c|c}
\toprule
Researchers &  GC~\cite{DBLP:conf/trec/GrossmanC17}  &  YXL~\cite{DBLP:conf/ecir/YuXL19} & BFFMSSS~\cite{DBLP:conf/sigir/Breuer0FMSSS20} \\
\midrule
\multicolumn{4}{c}{Baseline} \\
\midrule
Average Precision & 0.3711 &  0.4018  &  0.3612\\
Kendall's $\tau$  Union  & 1.0000  &  0.0086 &  0.0051\\
Rank-Biased Overlap  & 1.0000  &  0.1630 &  0.5747\\
Root Mean Square Error & 0.0000 &  0.1911 &  0.1071\\
p-value & 1.0000  &  0.1009 &  0.7885\\
\midrule
\multicolumn{4}{c}{Advanced} \\
\midrule
Average Precision   & 0.4278 &  0.4487 &  0.4208\\
Kendall's $\tau$  Union    & 1.0000  &  0.0069 &  0.0111\\
Rank-Biased Overlap    & 1.0000  &  0.2231 &  0.6706\\
Root Mean Square Error  & 0.0000  &  0.2088 &  0.0712\\
p-value  & 1.0000  &  0.2785 &  0.8249\\
\midrule
\multicolumn{4}{c}{Overall effects} \\
\midrule
Effect Ratio        & 1.0000 &  0.8267 &  1.0514 \\
$\Delta$ Relative Improvement  & 0.0000  &  0.0362 & -0.0123\\
\bottomrule
\end{tabularx}
\end{table}

As the AP scores show, for both run types (baseline and advanced), the retrieval performance of the BFFMSSS reimplementations is slightly below the original results, while the YXL reimplementations outperform the results by GC. We assume that the additional first-stage-ranker based on BM25 already provides a good baseline with acceptable recall rates which is of benefit for the ML reranker. 

Regarding the reproducibility, Kendall's $\tau$ Union shows that both reimplementations fail to reproduce the exact ordering of the documents in the rankings. Optimally, these scores should be close to $1.0$, while the reported values show that there is almost no correlation between the document rankings. However, in this regard, there is a higher similarity between the GC and reimplemented runs, especially for the BFFMSSS runs, in terms of the Rank-biased Overlap which can be partly explained by the measure's discount for lower ranks \cite{DBLP:journals/tois/WebberMZ10}.

Likewise, the RMSE and the p-values indicate that, in comparison, the topic score distributions of the BFFMSSS runs are closer to the GC runs than the reimplementations by YXL. Both measures are determined by the topic score distributions of the AP scores. While low p-values would result from different distributions, higher p-values indicate more similar distributions. As mentioned earlier, the YXL runs already result in strong baseline scores that outperform the original AP scores, and as a result, the p-value is lower. Regarding the advanced run, both reimplementations achieve slightly higher p-values.

If a baseline and advanced run are available, it is possible to determine the Effect Ratio (ER) and the $\Delta$ Relative Improvement (DRI). Both measures quantify how well the reimplementations preserve the improvements of the advanced run over the baseline. As shown by the ER, the BFFMSSS runs are closer to the optimal value of $1.0$, which again can be explained by the already strong baseline of YXL runs. As a result, the improvements of the advanced YXL runs are not as high as in the original experiments. Similarly, the DRI score, which also accounts for the absolute scores of the baseline runs, is closer to the optimal value.

\subsection{P'R'I'M'A'D': Generalization with other Data}
Finally, we provide an outlook on how well CCRF generalizes with other Data and with a Method based on training data from web-search results. This setup translates into a setting where every PRIMAD component is changed w.r.t. the original experiment. 

\begin{figure}
    \centering
    \includegraphics[width=\columnwidth]{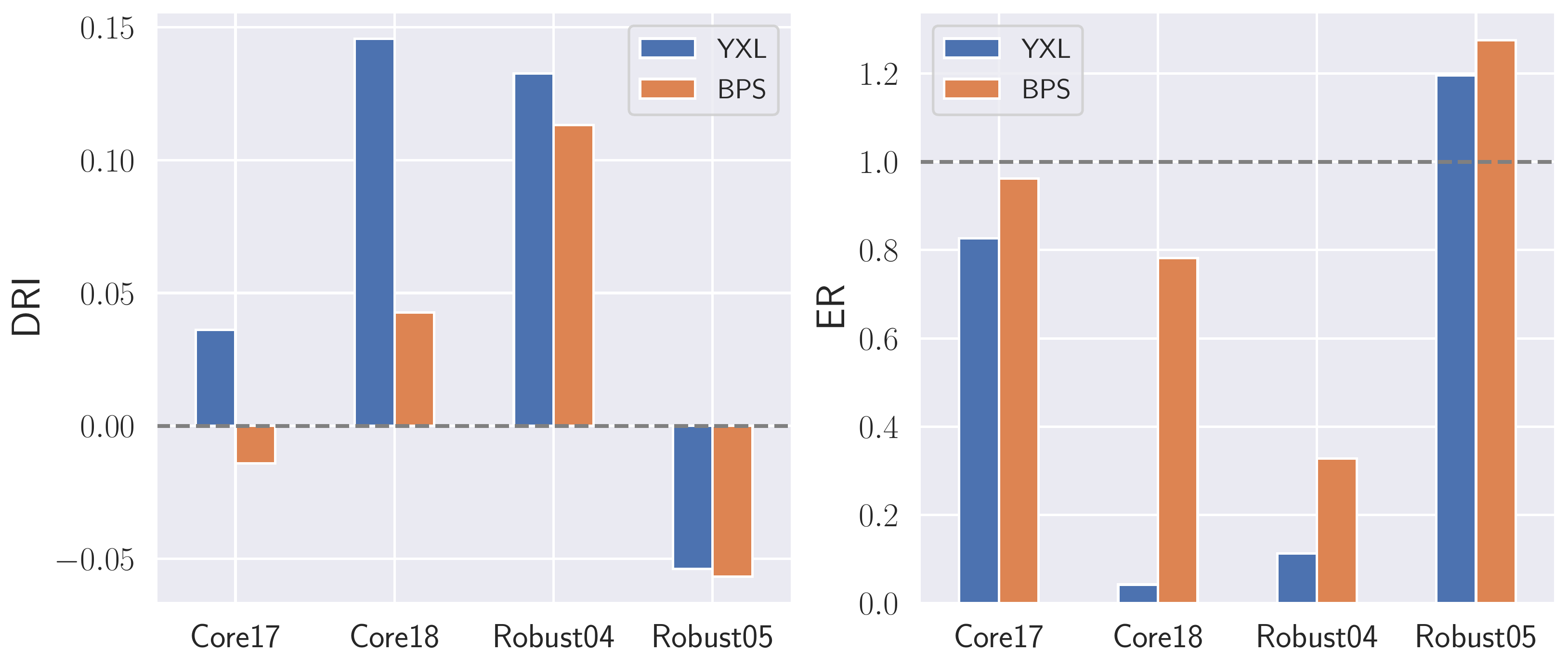}
    \caption{Overall effects measures by the Delta Relative Improvement (DRI) and the Effect Ratio (ER) of AP.}
    \label{fig:PRIMAD_er}
\end{figure}

From both reimplementations of YXL and BPS we evaluate runs derived from four target collections, including Core17/18 and Robust04/05. When reproducing a retrieval experiment on new Data, it is impossible to evaluate some of the reproducibility measures since they depend on runs derived for the same topics or from the same pool of documents. Instead, the previously introduced measures ER and DRI can be used as proxies to quantify how well relative improvements can be reproduced. For both run types, Figure \ref{fig:PRIMAD_er} illustrates these measures.

The YXL runs are evaluated with the GC runs from Core17 \cite{DBLP:conf/trec/GrossmanC17}, while the BPS runs are evaluated in comparison to the GC runs from Core18 \cite{DBLP:conf/trec/GrossmanC18}. Like in the previous Sections, we instantiate the measures with the AP scores. Both measures show that the reimplementations deviate from the original experiment except for the BPS runs of Core17 with a nearly optimal DRI value. Likewise, the corresponding ER score is below but close to 1.0. 

The YXL runs of Core18 have the largest DRI deviation and also the lowest ER score. This complies with the results of 6.1, which already showed that CCRF does not generalize well with this particular combination of the Robust corpora and Core18.

The experiments with Robust04 also show that CCRF does not generalize with this particular dataset. The positive DRI scores and low ER scores show that the baseline scores are higher than in the original experiment, while the reimplemented advanced runs do not provide a similar improvement. 

Both experiments on Robust05 have comparable ER scores above 1.0, which shows that the generalization was more successful than the experiments with the other test collections. However, it has to be considered that there are lower absolute scores, as shown by the negative DRI scores. 

Overall, we see that the reproducibility of CCRF strongly depends on the combinations of the datasets. While it is out of this study's scope to draw any conclusions about this circumstance, we assume this can be attributed to a higher overlap of vocabulary terms in documents with relevance assessments in the respective corpora. 

\section{Conclusions and outlook}
 
In this work, we present a novel metadata schema for TREC run files, one of the central experimental resources of IR research. In sum, PRIMAD is an adequate conceptual taxonomy for defining the schema, as shown by the literature review supporting the intentions of each component. Our annotated dataset of runs and the experiments exemplified what kinds of meta-evaluations are leveraged by comparing the PRIMAD metadata of annotated runs. 

By stimulating IR researchers and practitioners to annotate their runs with the outlined metadata schema, the robustness and reproducibility of IR (meta-)evaluations would be increased. Overall, designing and implementing a metadata schema is a socio-technological process that requires technical solutions and infrastructures, and the research community's adoption. In this regard, it is an important question how experimenters can be motivated to accept the additional overhead required for the annotations. 
As a starting point, our provided resources include the software support that allows automatic annotations for some of the PRIMAD components, which define the metadata schema. Furthermore, the project's website provides additional annotation help in the form of checklists for each component that can be used as a reference when annotating runs. As mentioned before, the website's source code is publicly hosted on GitHub, it is easy to maintain, and we aim at developing it in a collaborative process with the community in the future. By using common software development workflows, e.g., filing issues or making pull requests, it should be easy to adapt and extend the metadata schema while also paying attention to an adequate release versioning of the metadata schema.

With the help of the annotated dataset and the experiments, we could show that the schema can be applied to pipelines with ensembles and reranking stages fairly well. However, it needs to be elaborated on how the schema can be applied to more complex pipelines. There may be limitations to which degree the complexity of modern retrieval pipelines can be documented in metadata fields, and any further documentation must rely on the publication and the source code itself. On the other hand, it might be possible to parse certain retrieval pipelines from the source code itself into a more human-readable format. Thus, we would like to implement more software features in order to promote the adoption by the community even further. These features cover but are not limited to the following ideas, which also include and address the feedback by the anonymous reviewers of this work. 

While \texttt{repro\_eval} is a dedicated reproducibility toolkit and supports the outlined metadata schema, its features are limited to the (reactive) reproducibility analysis. Even wider adoption by the community could be reached if the schema is supported by commonly used retrieval toolkits like Pyserini or Pyterrier. For instance, the way Pyterrier defines the retrieval experiment in a declarative manner could be exploited to automatically extract the metadata annotations of the Method and make the runs proactively reproducible.

Currently, the metadata annotations follow the YAML syntax. We chose YAML because of its easy readability and extensibility. By using YAML, the annotations remain mostly free of markup artifacts, e.g., like there are known from XML-formatted data, making the annotations more human-readable. On the other hand, YAML is a recent and well-supported data-serialization language for which many well-curated parsing libraries exist. Its minimalistic syntax facilitates new metadata extensions while being both human- and machine-readable. We did not explicitly decide against following any existing metadata standard but preferred YAML because of its simplicity. In the future, it might be interesting to implement the support of existing standards or parse the \texttt{ir\_metadata} schema to other metadata formats. This would contribute to more sustainability of the introduced resource.

In order to provide more orientation and better feedback for metadata annotators, we are planning to develop software features that check the validity and integrity of the annotations. This feature could possibly use the already implemented automatic annotation features and give feedback on important missing metadata fields. To do this, the single metadata fields should be prioritized regarding their importance for reproducibility. Finally, we are planning to work more closely with shared task organizers who can enforce the annotations at submission time. In the long term, the metadata should become part of the run data archives to be used for adequate baselines and meta-evaluations.

\vspace{1.25em}
\small
\noindent \textbf{Acknowledgments}. This paper is supported by DFG (project no. 407518790).

\bibliographystyle{ACM-Reference-Format}
\bibliography{bibliography}

\end{document}